\documentclass[preprint, nofootinbib, floatfix, notitlepage]{revtex4-1}

\usepackage{amsmath,amsfonts,amssymb}
\usepackage{mathrsfs}
\usepackage{graphicx}
\usepackage[english]{babel} 
\usepackage{hyperref} 
\usepackage{color}

\usepackage{animate}

\usepackage[cal=boondoxo]{mathalfa}

\hypersetup{
    colorlinks=true,
    linkcolor=red,
    citecolor=blue,
} 

\newcommand{\be}{\begin{equation}}
\newcommand{\ee}{\end{equation}}
\newcommand{\mn}{{\mu \nu}}

\newcommand{\gy}{g_{Y}}

\begin{document}

\title{Gravitational Wave -- Gauge Field Dynamics}
\author{R. R. Caldwell}
\email{robert.r.caldwell@dartmouth.edu (corresponding author)}
\author{C. Devulder}
\email{christopher.devulder.gr@dartmouth.edu}
\affiliation{Department of Physics and Astronomy, Dartmouth College, 6127 Wilder Laboratory, Hanover, NH 03755 USA}
\author{N. A. Maksimova}
\email{nina.maksimova@cfa.harvard.edu}
\affiliation{Harvard-Smithsonian Center for Astrophysics, Cambridge, MA 02138, USA}
\date{\today}

\begin{abstract}

The dynamics of a gravitational wave propagating through a cosmic gauge field are dramatically different than in vacuum. We show that a gravitational wave acquires an effective mass, is birefringent, and its normal modes are a linear combination of gravitational waves and gauge field excitations, leading to the phenomenon of gravitational wave -- gauge field oscillations. These surprising results provide insight into gravitational phenomena and may suggest new approaches to a theory of quantum gravity. 

\end{abstract}
\maketitle

\noindent
Essay written for the Gravity Research Foundation 2017 Awards for Essays on Gravitation.

\vfill

\eject

General relativity is ever full of surprises. Over one hundred years since its inception \cite{Einstein:1915ca}, the frontiers of gravitation remain fertile subjects. The first detection of gravitational waves has established the field of gravitational wave astronomy \cite{Abbott:2016blz}. The search is on for the imprint of relic gravitational radiation of quantum origin on the polarization pattern of the cosmic microwave background \cite{Kamionkowski:2015yta}, and the accelerating cosmic expansion hints at quantum gravitational effects \cite{Riess:1998cb,Perlmutter:1998np}. In this milieu, cosmic gauge fields have been widely investigated for a possible role in catalyzing an inflationary epoch \cite{Maleknejad:2011jw,Adshead:2012kp,Adshead:2013qp,Namba:2013kia}. As we present in this essay, these investigations have revealed novel properties of the gravitational wave - gauge field system \cite{Caldwell:2016sut}. 

The key element in this work is a collection of non-Abelian gauge fields with a vacuum expectation value (vev). The action for the theory is
\be
	S = \int d^4 x \sqrt{-g} \left( \frac{M_P^2}{2}R-\frac{1}{4}F_{a\mn}F^{a \mn} \right),
	\label{eqn:thy}
\ee
where we use metric signature $-+++$ and $M_P$ is the reduced Planck mass. In the simplest realization, we consider an SU(2) gauge field with field strength tensor $F^{a}_\mn \equiv \partial_\mu A^{a}_\nu - \partial_\nu A^{a}_\mu - \gy \epsilon^{abc}A_{b\mu} A_{c\nu},$ where $\gy$ is the coupling constant. Greek letters represent space-time indices, and Latin letters $i,\,j,\, ...$ are spatial indices. The SU(2) indices are indicated by $a,\,b,\, ...=1,\,2,\,3$, and are raised and lowered by a metric $\eta = {\rm diag}(1,1,1)$.

We consider a configuration of the gauge field $A^a_\mu$ with stress-energy that is isotropic and homogeneous, consistent with the symmetries of the cosmological Robertson-Walker spacetime. In a coordinate system $ds^2 = a(\tau)^2(-d\tau^2 + d\vec x^2)$, we adopt the Ansatz $A_i^b = \phi (\tau) \delta^b_i$ with all other components vanishing.  In essence, we have identified the global part of the SU(2) with the O(3) rotational symmetry of spacetime. This ``flavor-space locked" field configuration  \cite{Bielefeld:2015daa} resembles a pair of uniform electric and magnetic fields for each flavor, pointing along the $x-,\, y-,\, z-$directions. Although the configuration is anisotropic in flavor, it is isotropic in pressure and energy. The equation of motion $\nabla^\mu F^{a}_{\mu\nu} + \gy \epsilon^{abc} A_b^\mu F_{c\mu\nu}=0$ reduces to $\phi'' + 2\gy^2 \phi^3 =0$ under our Ansatz, which is solved in terms of elliptic Jacobi functions. The classical field amplitude simply oscillates, and the flavor-space locked configuration under this model is stable, as shown through a linear perturbation analysis \cite{Dimastrogiovanni:2012ew,Bielefeld:2014nza}.

The gauge field strength tensor $F^{a}_{\mu\nu}$ has non-zero components where we expect to find an electric field, $``E" = F^a_{0i} = {\phi'}\delta^a_i/{a}$. Due to the coupling $\gy$ there is also a magnetic field, $``B"= F^a_{ij} = -\gy \phi^2 {\epsilon^a}_{ij}$, which, for each flavor, is coaligned with the electric field. This vev enables the gauge field to support transverse, traceless, synchronous tensor fluctuations which couple to gravitational waves.

In order to build intuition, we first consider a monochromatic gravitational wave propagating along the z-direction. We choose a circularly polarized gravitational wave, because the gauge field has a built-in right handedness in the group structure constants $\epsilon^{ijk}$. As it squeezes and stretches the gauge field along alternate axes in the $x-y$ plane, it enhances the $B$ field by an amount that is proportional to the gravitational wave amplitude, in phase with the wave. The $E$ field is also enhanced, but lags by $\pi/2$. These arguments suggest the $B$ and $E$ act like a spring and an anti-spring.

Second, we consider that the gauge field itself may fluctuate. By perturbing the full gauge field equation of motion we note that fluctuations of the field $A^a_\mu$ in each direction enhance the $E$ and $B$ energy, preferentially in a right-handed circular pattern. This, of course, reflects the built-in right-handedness. These rough arguments suggest that left- and right-circularly polarized fluctuations of the gauge field will propagate differently.

Third, by perturbing the gravitational field and the gauge field simultaneously, we deduce that the wave-like excitations couple. That is, a monochromatic gravitational wave can produce a wave-like excitation of the gauge field with the same wave number, and vice versa. This observation harks back to a remarkable series of papers starting with the work of Gertsenshteyn \cite{Gertsenshteyn1962}, in which the authors showed that a gravitational wave propagating through a stationary magnetic field converts into an electromagnetic wave and back again \cite{Poznanin1969,Boccaletti1970,Zeldovich1974}. In contrast to electromagnetism, the presence of three flavors allows us to build an isotropic medium.

Here we investigate the more general phenomenon of the conversion of a gravitational wave into a gauge field, as may be present in the early stages of the Universe. In particular, we discover that gravitational waves transform into gauge field waves, disappearing and reappearing much like neutrino flavor oscillations. 

We consider linearized gravitational waves and tensor fluctuations of the gauge field. Following a standard calculation, $\delta g_{ij} = a^2 h_P \, e^{P}_{ij}$ where $P=(L,\,R)$ labels the circular polarization and $e^{P}_{ij}$ is the standard polarization matrix. Similarly, we consider $\delta A^a_j = a\, y_P \, e^{P\, a}_j$ which makes use of the same polarization matrix. A change of variables, $h = H \sqrt{2} / {a M_P} $ and $y = Y/{\sqrt{2} a}$, puts the action into canonical form. The equations of motion, in terms of the Fourier amplitudes, are 
\begin{eqnarray}
&&H_L'' + \left[k^2 - \frac{a''}{a}  + \frac{2}{a^2 M_P^2}(\gy^2 \phi^4 - \phi'^2)\right] H_L = \frac{2}{a M_P}\left[ (\gy \phi + k) \gy  \phi^2 Y_L -  \phi' Y_L'\right]
\label{eqn:vL} \\
&&Y_L'' + \left[k^2 + 2 \gy k \phi \right]Y_L  
=\frac{2}{a M_P}\left[ a\left(\frac{\phi' }{a}H_L\right)' + \gy \phi^2 \left(k - \gy \phi  \right)  H_L  \right].
\label{eqn:uL}
\end{eqnarray}
The equations for $H_R,\,Y_R$ are obtained by replacing $k \to -k$. First, we observe, now quantitatively, that the gravitational wave equation acquires a time-dependent, mass-like term
\be
m^2 = \frac{2}{a^4 M_P^2}(\gy^2 \phi^4 - \phi'^2)
\ee
proportional to $B^2-E^2$, arising from the stress of the gauge field. Apart from the possibility that dominance of $E$ or $B$ can enhance or suppress a spectrum of long wavelength gravitational waves, there is a deeper point to be made. The effective mass term introduces a new scale into the system, and the gravitational wave amplitude is no longer comparable to a massless, minimally coupled scalar field \cite{Ford:1977dj}. This feature may suggest new approaches to the issues facing theories of massive gravity: the background matters. That is, a cosmological, spin-1 field may play an important role in a symmetry breaking scheme for the graviton.
  
Second, the coupled system is birefringent as revealed by the equations of motion for the left- and right-circular polarizations. In particular, the dispersion term $k^2 + 2 \gy k \phi$ can be negative for a range of wavenumbers; this holds for one polarization but not the other, thereby preferentially amplifying the gauge field and consequently the gravitational waves. This chiral asymmetry could have profound implications for the search for the imprint of primordial gravitational waves on CMB polarization anisotropy. It means there should be a unique, parity-odd correlation between temperature and the so-called ``B-modes" \cite{Kamionkowski:1996ks, Zaldarriaga:1996xe, Lue:1998mq}. This signal is already the target of current and planned experiments \cite{Ade:2015cao,Array:2015xqh, Abazajian:2016yjj, Kogut:2011xw, Bouchet:2011ck, Benson:2014qhw, Fraisse:2011xz, Suzuki:2015zzg}.

Third, at high frequencies the gravitational wave and gauge field interconvert through the phenomenon of gravitational wave -- gauge field oscillations \cite{Caldwell:2016sut}.  This is more easily seen if we focus on the Lagrangian for gravitational and gauge field waves propagating with Fourier wavenumber $k$ greater than the expansion rate or rate of change of the gauge field,  
\begin{equation}
{\cal L} = \frac{1}{2}H_L'^2 - \frac{1}{2}k^2 H_L^2 + \sum_{n=1}^{\cal N}\left[ \frac{1}{2}Y_{Ln}'^2 - \frac{1}{2}k^2 Y_{Ln}^2- k \gy \phi Y_{Ln}^2 + \frac{2}{a M_P}H_L\left(k \gy\phi^2Y_{Ln} - \phi' Y_{Ln}'\right)\right].\,\,\,
\label{eqn:LHY}
\end{equation}
We have now allowed ${\cal N}$ families of SU(2), embedded in a larger SU(N), where ${\cal N}=[N/2]$. At the most basic level, the Lagrangian above describes ${\cal N}+1$ coupled oscillators. At high frequency, $H$ and each of $Y_n$ oscillate with frequency $k$. The gravitational wave couples to each gauge field wave; each gauge field wave couples only to the gravitational wave. 

\begin{figure}[t]
\includegraphics[width=0.45\textwidth,angle=0]{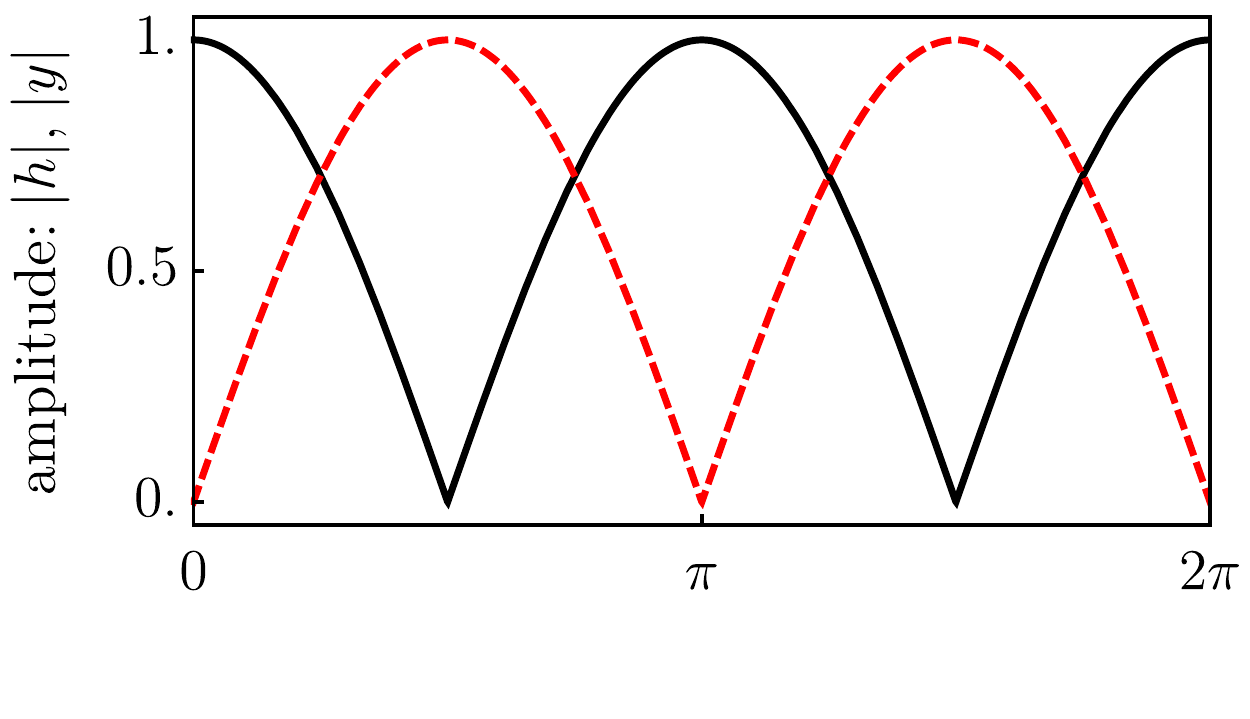}
\includegraphics[width=0.45\textwidth,angle=0]{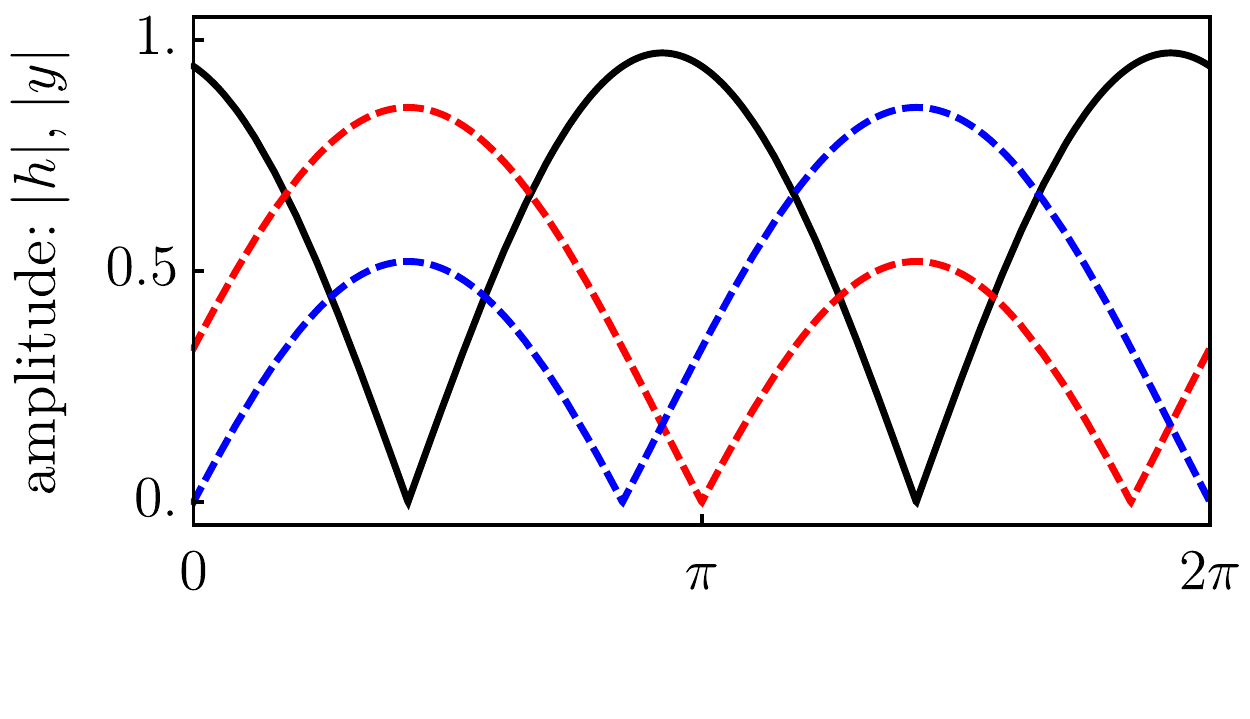}
\vspace{-0.75cm}
\includegraphics[width=0.45\textwidth,angle=0]{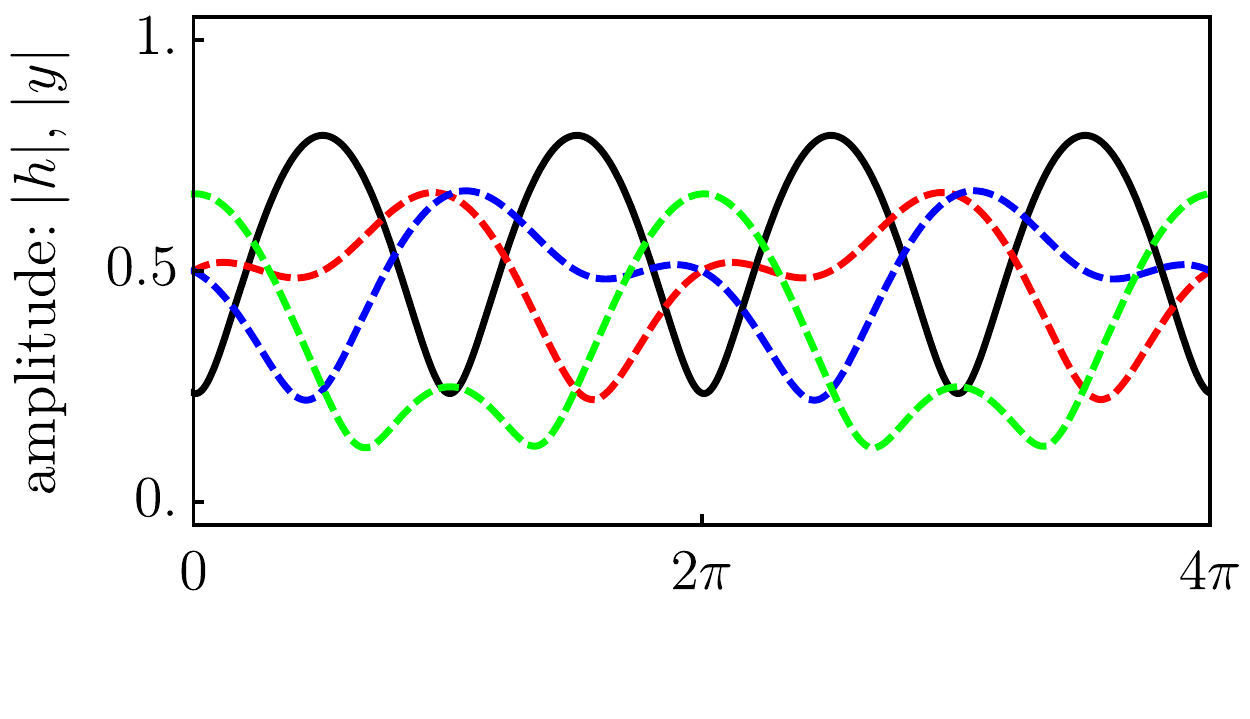}
\includegraphics[width=0.45\textwidth,angle=0]{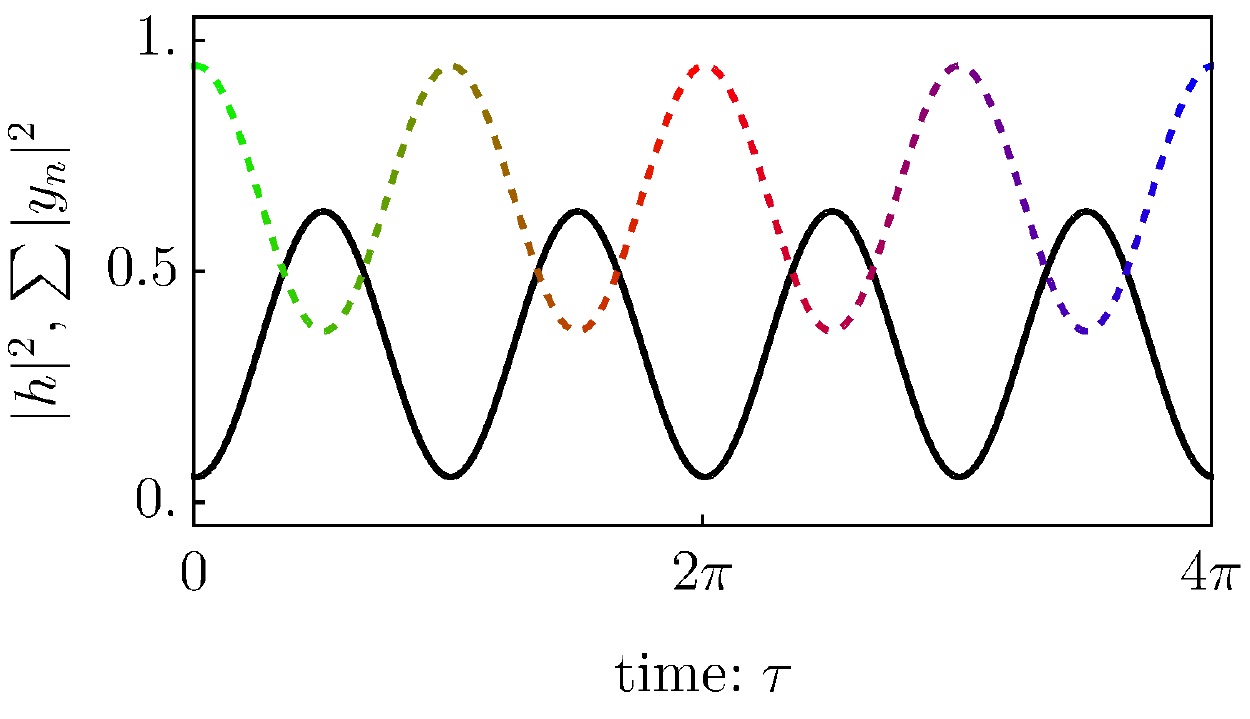}
\caption{The oscillations of the gravitational wave amplitude ${\cal h}$ (black) and gauge field ${\cal y}$ (dashed) are shown for ${\cal N}=1,\,2$ and $3$. In all cases, $|\psi|^2=1$ and the oscillation period has been scaled. The bottom right panel shows that the squared amplitudes sum to unity. (Figures reproduced from Ref.~\cite{Caldwell:2016sut}.)}
\label{fig:fig1}
\end{figure}

The {\it gravitational wave -- gauge field oscillations} are revealed by the rms amplitude of the waves in the high frequency limit. We write $H = {\cal h} e^{-i k \tau}$ and $Y_n = {\cal y}_n e^{-i k \tau}$ and choose $k$ to be sufficiently large such that we can treat the coefficients $\phi,\, \phi'$ as constants. The normal modes are identified by diagonalizing the Lagrangian (\ref{eqn:LHY}). Starting with the gravitational and gauge field modes $\psi^i = ({\cal h},\, {\cal y}_n)$ for $i=0,\,1, ...\,,\, {\cal N}$, we write the Lagrangian in the form ${\cal L} = \tfrac{1}{2}\psi'^\dagger \mathbb{I} \psi' - \tfrac{1}{2}\psi^\dagger M^2 \psi$. We transform into the eigenbasis of $M^2$ via $\psi^i = R^i_j \Delta^j$, where $\Delta^j= (\Delta_0,\, \Delta_n)$ are the normal modes. Hence, the Lagrangian acquires the form ${\cal L} = \tfrac{1}{2}\Delta'^\dagger \mathbb{I} \Delta' - \tfrac{1}{2}\Delta^\dagger \Omega^2 \Delta$ and $\Omega^2$ is diagonal with the normal mode frequencies. 

Conservation of the canonical stress-energy tensor $\Theta^{\mu\nu} = \partial^\mu \psi^i \delta {\cal L}/\delta \partial_\nu \psi^i - \eta^{\mu\nu} {\cal L}$ yields the constant of motion in the high frequency limit, $|\psi|^2 = |{\cal h}|^2 + \sum_{n=1}^{\cal N} |{\cal y}_n|^2 = \sum_{n=0}^{\cal N}|c_n|^2,$ where the coefficients $c$ are the initial values of ${\cal h},\,{\cal y}_n$. Hence, we determine that the gravitational and gauge field wave amplitudes trace a pattern on the surface of an ${\cal N}+1$-dimensional sphere, endowing the system with an emergent symmetry. This behavior is reminiscent of neutrino flavor oscillations, where the mass eigenstates remain in phase while the flavor eigenstates oscillate. Examples of the oscillation patterns are shown in Fig.~\ref{fig:fig1} and as an animation in Fig.~\ref{fig:fig4}.

\begin{figure}[t]
\begin{frame}{}
\animategraphics[loop,controls,width=0.4\linewidth]{12}{GWGFO60-}{0}{60}
\end{frame} 
\caption{The gravitational wave -- gauge field oscillations for the case ${\cal N}=3$ are illustrated using oscillating pistons. The gravitational and gauge field waves are represented by the central and surrounding pistons. (Figure reproduced from Ref.~\cite{Caldwell:2016sut}. View in Adobe Reader to play animation.)}
\label{fig:fig4}
\end{figure}

The gravitational wave -- gauge field oscillations have implications for the quantization of the gravitational field. For example, in a gauge-field inflationary scenario, the Hilbert space must be expanded to include the gauge field excitations. Quantum fluctuations in the gravitational field give rise to a homogeneous solution $H_h$, whereas fluctuations in the gauge fields create $n=1, ..., {\cal N}$ inhomogeneous solutions $H_{i,\,n}$. The power spectrum is $\langle H^2\rangle = |H_h|^2 + \sum_{n=1}^{\cal N} |H_{i,\,n}|^2$, which reflects the new emergent symmetry. The modulation of the individual amplitudes now cancels, yielding a constant amplitude $|\psi|^2$ which is reflected in the observable power spectrum.

The example investigated in this essay has afforded us a deeper understanding of gravity. The gauge field introduces a mass scale and breaks parity in the gravitational wave system. A new symmetry also emerges among the gravitational wave and gauge field amplitudes. These effects may have a unique imprint on a spectrum of primordial gravitational waves, allowing us to gain unparalleled insight into the earliest moments in the history of the Universe. Moreover, these results suggest that to go beyond linear order in the quantization of gravity, the expanded set of gravitational degrees of freedom created by the background must be taken into account.

\vfill
\acknowledgments
The work of RRC and CD is supported in part by DOE grant DE-SC0010386. The work of NAM is supported by the National Science Foundation Graduate Research Fellowship under Grant No. DGE1144152.
 
\vfill



\begin{thebibliography}{99}

\bibitem{Einstein:1915ca} 
  A.~Einstein,
  Sitzungsber.\ Preuss.\ Akad.\ Wiss.\ Berlin (Math.\ Phys.\ ) {\bf 1915}, 844 (1915).
  
\bibitem{Abbott:2016blz} 
  B.~P.~Abbott {\it et al.} [LIGO Scientific and Virgo Collaborations],
  Phys.\ Rev.\ Lett.\  {\bf 116}, no. 6, 061102 (2016)
  doi:10.1103/PhysRevLett.116.061102
  [arXiv:1602.03837 [gr-qc]].

\bibitem{Kamionkowski:2015yta} 
  M.~Kamionkowski and E.~D.~Kovetz,
  Ann.\ Rev.\ Astron.\ Astrophys.\  {\bf 54}, 227 (2016)
  doi:10.1146/annurev-astro-081915-023433
  [arXiv:1510.06042 [astro-ph.CO]].
  
\bibitem{Riess:1998cb} 
  A.~G.~Riess {\it et al.} [Supernova Search Team],
  Astron.\ J.\  {\bf 116}, 1009 (1998)
  doi:10.1086/300499
  [astro-ph/9805201].

\bibitem{Perlmutter:1998np} 
  S.~Perlmutter {\it et al.} [Supernova Cosmology Project Collaboration],
  Astrophys.\ J.\  {\bf 517}, 565 (1999)
  doi:10.1086/307221
  [astro-ph/9812133].

\bibitem{Maleknejad:2011jw} 
  A.~Maleknejad and M.~M.~Sheikh-Jabbari,
  Phys.\ Lett.\ B {\bf 723}, 224 (2013)
  [arXiv:1102.1513 [hep-ph]].

\bibitem{Adshead:2012kp} 
  P.~Adshead and M.~Wyman,
  Phys.\ Rev.\ Lett.\  {\bf 108}, 261302 (2012)
  doi:10.1103/PhysRevLett.108.261302
  [arXiv:1202.2366 [hep-th]].

\bibitem{Adshead:2013qp} 
  P.~Adshead, E.~Martinec and M.~Wyman,
  Phys.\ Rev.\ D {\bf 88}, no. 2, 021302 (2013)
  doi:10.1103/PhysRevD.88.021302
  [arXiv:1301.2598 [hep-th]].

\bibitem{Namba:2013kia} 
  R.~Namba, E.~Dimastrogiovanni and M.~Peloso,
  JCAP {\bf 1311}, 045 (2013)
  doi:10.1088/1475-7516/2013/11/045
  [arXiv:1308.1366 [astro-ph.CO]].
 
\bibitem{Caldwell:2016sut} 
  R.~R.~Caldwell, C.~Devulder and N.~A.~Maksimova,
  Phys.\ Rev.\ D {\bf 94}, no. 6, 063005 (2016)
  doi:10.1103/PhysRevD.94.063005
  [arXiv:1604.08939 [gr-qc]].
 
\bibitem{Bielefeld:2015daa} 
  J.~Bielefeld and R.~R.~Caldwell,
  Phys.\ Rev.\ D {\bf 91}, no. 12, 124004 (2015)
  doi:10.1103/PhysRevD.91.124004
  [arXiv:1503.05222 [gr-qc]].
 
\bibitem{Bielefeld:2014nza} 
  J.~Bielefeld and R.~R.~Caldwell,
  Phys.\ Rev.\ D {\bf 91}, no. 12, 123501 (2015)
  doi:10.1103/PhysRevD.91.123501
  [arXiv:1412.6104 [astro-ph.CO]].

\bibitem{Dimastrogiovanni:2012ew} 
  E.~Dimastrogiovanni and M.~Peloso,
  Phys.\ Rev.\ D {\bf 87}, no. 10, 103501 (2013)
  doi:10.1103/PhysRevD.87.103501
  [arXiv:1212.5184 [astro-ph.CO]].
   
\bibitem{Gertsenshteyn1962}
M. E. Gertsenshteyn,
Sov.\ Phys.\ JETP {\bf 14}, 84 (1962); Zh. Eksp. Teor. Fiz. (U.S.S.R.) {\bf 41}, 113 (1961).

\bibitem{Poznanin1969}
P.-L. Poznanin,
Sov.\ Phys.\ J.\ {\bf 12}, 1296 (1969), doi:10.1007/BF00815672.

\bibitem{Boccaletti1970}
D. Boccaletti, V. Sabbata, P. Fortini, and C. Gualdi,
Nuovo Cimento V Serie {\bf 70}, 129 (1970), doi:10.1007/BF02710177.

\bibitem{Zeldovich1974}
Ya. B. Zeldovich,
Sov.\ Phys.\ JETP {\bf 38}, 652 (1974); Zh. Eksp. Teor. Fiz. (U.S.S.R.) {\bf 65}, 1311 (1973).

\bibitem{Ford:1977dj} 
  L.~H.~Ford and L.~Parker,
  Phys.\ Rev.\ D {\bf 16}, 1601 (1977).
  doi:10.1103/PhysRevD.16.1601
   
\bibitem{Kamionkowski:1996ks} 
  M.~Kamionkowski, A.~Kosowsky and A.~Stebbins,
  Phys.\ Rev.\ D {\bf 55}, 7368 (1997)
  doi:10.1103/PhysRevD.55.7368
  [astro-ph/9611125].

\bibitem{Zaldarriaga:1996xe} 
  M.~Zaldarriaga and U.~Seljak,
  Phys.\ Rev.\ D {\bf 55}, 1830 (1997)
  doi:10.1103/PhysRevD.55.1830
  [astro-ph/9609170].

\bibitem{Lue:1998mq} 
  A.~Lue, L.~M.~Wang and M.~Kamionkowski,
  Phys.\ Rev.\ Lett.\  {\bf 83}, 1506 (1999)
  doi:10.1103/PhysRevLett.83.1506
  [astro-ph/9812088].
  
\bibitem{Ade:2015cao} 
  P.~A.~R.~Ade {\it et al.} [POLARBEAR Collaboration],
  Phys.\ Rev.\ D {\bf 92}, 123509 (2015)
  doi:10.1103/PhysRevD.92.123509
  [arXiv:1509.02461 [astro-ph.CO]].
    
\bibitem{Array:2015xqh} 
  P.~A.~R.~Ade {\it et al.} [BICEP2 and Keck Array Collaborations],
  Phys.\ Rev.\ Lett.\  {\bf 116}, 031302 (2016)
  doi:10.1103/PhysRevLett.116.031302
  [arXiv:1510.09217 [astro-ph.CO]].
  
\bibitem{Abazajian:2016yjj} 
  K.~N.~Abazajian {\it et al.} [CMB-S4 Collaboration],
  arXiv:1610.02743 [astro-ph.CO].
  
\bibitem{Kogut:2011xw} 
  A.~Kogut {\it et al.},
  JCAP {\bf 1107}, 025 (2011)
  doi:10.1088/1475-7516/2011/07/025
  [arXiv:1105.2044 [astro-ph.CO]].
  
\bibitem{Bouchet:2011ck} 
  F.~R.~Bouchet {\it et al.} [COrE Collaboration],
  arXiv:1102.2181 [astro-ph.CO].
  
\bibitem{Benson:2014qhw} 
  B.~A.~Benson {\it et al.} [SPT-3G Collaboration],
  Proc.\ SPIE Int.\ Soc.\ Opt.\ Eng.\  {\bf 9153}, 91531P (2014)
  doi:10.1117/12.2057305
  [arXiv:1407.2973 [astro-ph.IM]].
  
\bibitem{Fraisse:2011xz} 
  A.~A.~Fraisse {\it et al.},
  JCAP {\bf 1304}, 047 (2013)
  doi:10.1088/1475-7516/2013/04/047
  [arXiv:1106.3087 [astro-ph.CO]].
  
\bibitem{Suzuki:2015zzg} 
  A.~Suzuki {\it et al.} [POLARBEAR Collaboration],
  J.\ Low.\ Temp.\ Phys.\  {\bf 184}, no. 3-4, 805 (2016)
  doi:10.1007/s10909-015-1425-4
  [arXiv:1512.07299 [astro-ph.IM]].
  
 
\end{thebibliography}
\end{document}